# Handwriting Imagery EEG Classification based on Convolutional Neural Networks

Hao Yang[1], Guang Ouyang[2]
[1]Centre for Nonlinear Studies, Hong Kong Baptist University
[2]Complex Neural Signals Decoding Lab, Faculty of Education, The University of Hong Kong
Corresponding author: Guang Ouyang, email: ouyangg@hku.hk

**Abstract**
Handwriting imagery has emerged as a promising paradigm for brain-computer interfaces (BCIs) aimed at translating brain activity into text output. Compared with invasively recorded electroencephalography (EEG), non-invasive recording offers a more practical and feasible approach to capturing brain signals for BCI. This study explores the limit of decoding non-invasive EEG associated with handwriting imagery into English letters using deep neural networks. To this end, five participants were instructed to imagine writing the 26 English letters with their EEG being recorded from the scalp. A measurement of EEG similarity across letters was conducted to investigate letter-specific patterns in the dataset. Subsequently, four convolutional neural network (CNN) models were trained for EEG classification. Descriptively, the EEG data clearly exhibited letter-specific patterns serving as a proof-of-concept for EEG-to-text translation. Under the chance level of accuracy at 3.85%, the CNN classifiers trained on each participant reached the highest limit of around 20%. This study marks the first attempt to decode non-invasive EEG associated with handwriting imagery. Although the achieved accuracy is not sufficient for a usable brain-to-text BCI, the model's performance is noteworthy in revealing the potential for translating non-invasively recorded brain signals into text outputs and establishing a baseline for future research.

## 1. Introduction

Brain-computer interfaces (BCIs) have emerged as a cutting-edge neural technology that enables direct communication between the brain and external devices. One of the key components in BCI is the classification of brain signals using machine learning methods – in recent times – deep neural networks. Electroencephalography (EEG) is the dominant brain signal acquisition technology owing to its ability to capture high temporal resolution neural dynamic activity to cope with the need for BCI. Since EEG signals are typically stored in a two-dimensional array with one dimension for electrodes and another one time points, EEG classification with Convolutional Neural Networks (CNNs) has been intensively studied. Various CNN architectures have shown promising results in decoding specific brain patterns for basic cognitive research or for translating them into meaningful commands or actions [8][17][18].

Among the extensive research on CNNs for EEG classification, several prominent convolutional module designs have been proposed and demonstrated success across various BCI domains. Since the crucial information of EEG signals lies in the temporal dynamics of brain activities distributed across electrodes, temporal and spatial convolution have become common practices in neural network architectures for EEG classification. Schirrmeister et al. [20] investigated the performance of ShallowConvNet and DeepConvNet in EEG decoding, both of which perform a temporal convolution followed by a spatial filter in the first two layers. Lawhern et al. [6] proposed the well-known EEGNet, which utilizes a depthwise convolution for spatial filtering followed by a separable convolution. The EEGNet has served as the basis for various extensions, such as the EEG-CDILNet by Liang et al. [7] and the EEG-TCNet by Ingolfsson et al. [4]. More recently, the concept of Inception originally introduced by Szegedy et al. [22] for image classification has also been adapted for EEG decoding. The Inception module executes convolutions of different sizes in parallel to extract features at multiple temporal scales, which has been shown to be effective for EEG decoding tasks [18][19].

Besides the different forms of convolution mentioned above, the attention mechanisms first proposed by Bahdanau et al. [2] have also been incorporated into CNNs for EEG classification. The attention mechanisms allow the neural network to selectively focus on specific regions in the spatiotemporal space of EEG, such as certain electrodes and time points, by generating attention scores from given inputs and then processing the inputs in a way that incorporates the attention scores. For examples, Song et al. [21] employed a self-attention layer following a temporal and a spatial convolutional layer, and Miao et al. [9] incorporated a channel-attention module into a baseline model which combines DeepConvNet and EEGNet. As a more comprehensive study, Wimpff et al. [25] explored various attention mechanisms incorporated into a calibration model. In summary, CNN is hitherto the core architecture for processing the 2-D (electrodes and time points) EEG data with or without incorporating other modules for enhancing the extraction of relevant information.

In addition to the decoding algorithm variety, there is also a high degree of variety in the BCI paradigms. Brain-to-text BCI is a novel type of BCI with increasing popularity over the recent years [1][16]. It aims to convert brain activity into text or speech, which provides a potential means of communication for individuals with verbal disabilities. There are various types of BCI paradigms that enable the conversion of brain activity into texts. In a brain-to-text BCI based on visually evoked potentials (VEPs), for example, steady-state visually evoked potentials (SSVEPs), characters are displayed on the screen and flicker at unique frequencies, and the brain response to the flickering tells us which character the participant is focusing on [3][10][27]. In a brain-to-text BCI based on P300 waves, characters are presented in an array with multiple rows and columns. Different characters flash asynchronously at a frequency much lower than that is used in SSVEP. The character to which the users pay attention can be identified based on the amplitude of the associated P300 [1][11][15]. In a motor-imagery-based brain-to-text BCI, characters are displayed on a computer screen, and users imagine performing movements in a specific direction (similar to controlling a mouse cursor), and the corresponding brain signals are decoded to indicate the characters towards which the users want to move [1][5][11].

These existing BCI paradigms have made remarkable progress in terms of classification accuracy and decoding efficiency. However, they do not offer a direct conversion of imagery-based brain signals into texts. Instead, they reply on external information inputs, which is not naturalistic, and some of them require sustained attention to external tasks. For instance, SSVEP- and P300-based BCIs involve prolonged exposure to flickering stimuli, which can easily lead to fatigue in the user. Owing to these limitations, decoding handwriting imagery has attracted more and more attention because of its potential for developing more direct and naturalistic brain-to-text BCIs. Handwriting imagery entails neural processes for imagining intricate trajectories, which implies that the brain signals encoding handwriting may include richer temporal information for classification. Willett et al. [24] theorized that neural patterns associated with temporally complex trajectories are easier to classify than those linked to straight-line movements. They also showed that increased temporal dimensionality can make neural patterns of movements more distinguishable. Consequently, handwriting imagery holds promise as a more direct brain-to-text BCI paradigm that is inherently easier to decode.

Invasive recording techniques have demonstrated unprecedented success in decoding handwriting imagery in recent years. For instance, Willett et al. [24] developed an intracortical BCI that decodes handwriting from neural activity recorded by an implanted microelectrode array, which enables a patient with paralysis to type characters at a remarkable speed of 90 characters per minute [24]. However, invasive recording techniques are currently limited to patients undergoing craniotomy. As a remedy, attempts have also been made to decode handwriting imagery using non-invasive scalp recording, which offers advantages in terms of accessibility and usability [14][23]. Pei & Ouyang [14] examined the information related to handwriting encoded in non-invasive scalp-recorded brain signals to see whether it is feasible for classification. The accuracy they reached was impressively high (>90%). However, the task they used was based on the actual writing of the characters, which is not a true brain-to-text BCI. In sum, probably due to the various sources of noise in scalp-recorded signals, the application of non-invasive recording to brain-to-text BCIs is relatively unexplored, and the potential to translate scalp-recoded signals to effective communication remains unknown.

In this paper, we explore the potential of using scalp-recoded EEG associated with handwriting imagery to develop brain-to-text BCI. Specifically, we investigate the potential of using convolutional neural networks to decode single English letters. To the best of our knowledge, this work is the pioneering endeavor in decoding non-invasively recorded EEG associated with handwriting imagery. We aim to assess the potential performance of the models, specifically, the highest classification accuracy achievable by the neural networks we employed. By exploring this novel paradigm for a brain-to-text BCI, we seek to confirm the feasibility of non-invasive EEG of handwriting imagery for efficient communication and establish a baseline for future research. In the following sections, we will first introduce the dataset used in this study and then outline the preprocessing methods. Next, we will describe the four neural network models employed in this study, including the considerations and rationales behind the choice of the network architectures and hyperparameters. Finally, we will present relevant results and discussions on the study.

## 2. Methods
### 2.1 Participants and dataset

Our dataset consists of EEG signals from five participants. This study was approved by the Human Research Ethics Committee (HREC), The University of Hong Kong, under the approval number EA2006007. All methods were performed in accordance with the relevant guidelines, specifically regarding the protection of privacy and the right to withdraw from the study. All the participants provided written informed consent to participate in the experiment. All participants were healthy, right-handed individuals with no history of neurological or psychiatric disorders.

The participants were instructed to imagine handwriting capitalized English letter in each session. Because machine learning typically requires a large sample size to train and EEG experiments typically generate very few samples (trials), we separated the participant's EEG recording into 10 sessions on different days. Each participant visited the lab 10 times, generating 5×10 = 50 sets of EEG data in the conventional sense. Each session lasted for around 1 hour. In each session, the participants imagined writing 780 English letters, which was further divided into 10 blocks (to allow for intermittent breaks). In one block (Fig. 1), each of the 26 English letters was presented three times, generating 78 letters ordered randomly. The participant sat in front of a computer screen and wore an EEG cap recording the activity on the scalp. Each letter was shown on the screen for 200 ms and disappeared. In the following 3 seconds, the participant was required to imagine the process of handwriting this letter.

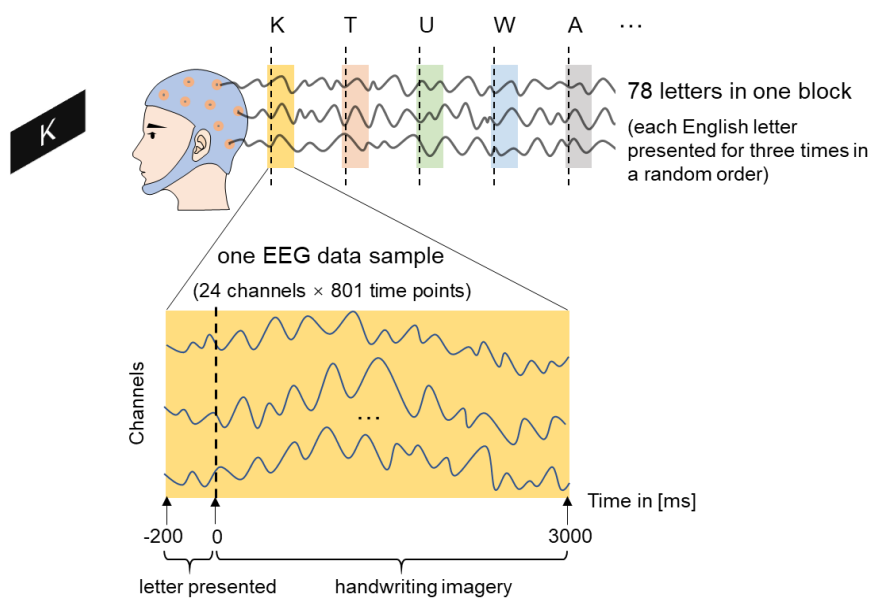

Fig. 1. Illustration of the handwriting imagery task. The participant sat in front of a computer screen and wore an EEG cap recording the brain activity. In one block, each of the 26 English letters was presented three times, resulting in a total number of 78 trials presented in random order. Each letter was shown on the computer screen for 200 ms and then the participants imagined writing that letter in the following 3 seconds. The participants completed 10 blocks in one session which was repeated for 10 times separated into 10 visits to the lab. Therefore, we obtained 78×10×10 = 7800 samples in total, with 3×10×10 = 300 samples for each English letter.

To reduce variability in the participants' imagery, they were required to imagine writing each letter stroke-by-stroke, following a standardized stroke order of common handwriting convections. This ensured that each imagined letter adhered to a consistent movement sequence across different participants. Participants were also asked to maintain a consistent spatial direction of each stroke for the same letter, and the size of the characters should match typical handwriting. Before the main experiment, participants underwent a training session where they were guided to imagine a few examples of simple and complex letters, ensuring that they understood the stroke order and spatial layout of the characters as well as got comfortable with the EEG device. In the main experiment, the participants were encouraged to focus on the trajectory of the imagined characters and to avoid overt motor actions, including moving body or stretching arms, hands, or legs.

The dataset is organized in a typical machine-learning setting. Each participant has completed 300 samples for each letter, so the dataset has 300×26 = 7800 samples in total. The cap has 24 electrodes and the sampling frequency is 250 Hz, so each sample has the shape of 24×801, where 24 is the number of channels and 801 is the number of time points. The time sequence starts at 200 ms before the letter is present and end at 3000 ms when the time is up for handwriting imagery.

**2.2. Data preprocessing**

Common preprocessing steps for EEG signals were performed on the data, including digital filtering, artifact removal, epoching, and baseline adjustment. The choice of frequency range of the filters and time range of the epoch were partly based on empirical knowledge and partly exploratory as the primary aim of the current study is to evaluate the upper limit of imagery-based handwriting classification from non-invasive EEG. A bandpass filter was applied to the original EEG data to remove the high-frequency noise and low-frequency shift. The frequency range was chosen to be 0.1 to 45 Hz which is commonly used in EEG signal processing. The data were then epoched from 0 to 1600 ms after the stimulus onset. Common artifacts in EEG, including large-amplitude artifacts caused by muscle activity and artifacts caused by eye movement, were removed using independent component analysis (ICA) methods.

In addition, we normalized the data before feeding them into the neural network to increase the stability of the training process. To preserve the spatial structure for downstream classification, for each sample, we first flatten the data into a one-dimensional vector, concatenating the time series of each channel. Next, we apply z-score normalization on the flattened data, subtracting each data point by their mean value and then dividing them by their standard deviation. Finally, we reshape the data back to 2 dimensions (channels by time points).

**2.3. Descriptive analysis**

Before conducting the machine learning approach for the decoding purpose, we first conducted some descriptive analysis to evaluate the feasibility of machine learning classification and also provide readers with some intuitive insights about our dataset. The main descriptive analysis we conducted was to evaluate the within-letter similarity and cross-letter dissimilarity in the EEG segment patterns. Specifically, we expected that two EEG patterns would be more similar in a certain way if they are from the same letter (e.g., A and A), and more dissimilar if they are from different letters (e.g., A and B). Being able to characterize this feature would provide solid evidence that machine learning would work (although not necessarily reaching a high accuracy). The similarity is measured by Pearson correlation coefficients of the averaged signals from two letters. The method of calculation is described as the following. We split the 300 samples of each letter into two halves and computed their average in each half. Thus, each letter generates two average EEG patterns, and the 26 letters generate 52 average patterns in total. We then calculate the pair-wise similarity of these 52 patterns, expecting that same-letter similarity is higher than different-letter similarity. This will generate a similarity matrix with a visible diagonal line (the values on the diagonal line are higher than non-diagonal ones). Note that, here, the expectation of higher values on the diagonal is non-trivial because the two average patterns of the same letter for similarity calculation are not from the same data source. The similarity matrix is calculated for each channel (electrode), further averaged across the channels. We calculate the similarity matrix following these steps for each subject. We further investigate the similarity pattern under different frequency components and different time windows based on a representative subject. The results will be presented in Section 3.1.

To better elucidate the letter-specific patterns in our dataset, we visualized the average signals of two contrasting English letters. We selected two letters to contrast based on their handwriting trajectory differences which are expected to generate different neural dynamics [24]. For instance, G and I are expected to undergo very different handwriting dynamics in the neural cognitive activity. Since EEG patterns contain both temporal and spatial information, we expected the same-letter similarity and different-letter dissimilarity to exist in both the temporal dynamics and scalp distribution of the average signals. To capture the temporal characteristics, we conducted principal component analysis (PCA) on the average ERP patterns using the scalp recordings at each time point as observations. This allows us to extract the principal components that best explain the temporal dynamics of the average signals. By visualizing the first three principal components, we will be able to reveal the temporal characteristics and how they account for the within-letter similarity and between-letter dissimilarity. Additionally, we visualized the scalp distributions of the potentials at selected time points to assess the spatial properties of the average signals. This combined visualization of temporal and spatial patterns was intended to provide a comprehensive picture of the letter-specific patterns in our EEG dataset.

### 2.4. Neural Network Architecture

To thoroughly investigate the potential of our scalp-recorded EEG dataset for handwriting imagery decoding, we employed diverse state-of-the-art convolutional neural network architectures in this study. Specifically, we utilized the following four models: the DeepConvNet by Schirrmeister et al. [20], the EEGNet by Lawhern et al. [6], the EEGInception by Santamaría-Vázquez et al. [19], and the LMDA by Miao et al. [9]. Based on our understanding of the constituent modules in these networks as well as our insights into the characteristics of mental activities underlying handwriting, we have made considerable adjustments to the original architectures to make them compatible with our dataset. Additionally, we performed hyperparameter fine-tuning on these models to optimize their classification capabilities for our decoding task.

The four network architectures use different convolutional layers for EEG decoding. The DeepConvNet represents a typical design which employs temporal convolution to extract dynamical patterns across all channels, followed by spatial filtering to integrate the temporal information while squeezing out the channel dimension. The resulting feature maps are further processed by sequential convolutional blocks to extract deeper representations for final classification. The EEGNet employs a more compact design, performing separable spatial convolution on the outputs of the temporal convolution by applying distinct spatial filters for each temporal feature. The EEGInception utilizes Inception modules to extract information at multiple temporal scales, with two Inception modules followed by two sequential convolutional blocks and the classification head. Each Inception module conducts three independent temporal convolutions of different sizes and then concatenates the three feature outputs. Lastly, the LMDA incorporates attention mechanisms, including a channel attention module that maps channel information to the depth dimension, and a depth attention module which enhances the interactions between different depths in the feature maps. The depth attention module acts between a temporal and a spatial convolution block, while the channel attention module is positioned before the temporal convolution block. Across all these models, the convolution blocks are interspersed with Batch Normalization, Dropout, Activation, and Pooling to increase the stability and generalizability of the model.

In order to adapt to the current EEG data features to explore the highest potential of different networks, we have made adjustments to the network architectures and their hyperparameters for our EEG classification. There are primarily two considerations guiding these modifications. Firstly, our dataset has a relatively limited sample size (300 samples per class) compared to typical deep learning tasks like image classification. Therefore, deploying an excessively deep neural network architecture would likely lead to overfitting issues. Consequently, for the DeepConvNet, we reduced the number of sequential convolutional blocks from 4 to 2 to better fit with our decoding purposes.

Secondly, since handwriting is an intricate motor skill involving fine control [12][13], we expect that handwriting imagery would involve subtle dynamics in brain activities occurring at a very short time scale, at the order of tens of milliseconds. To better capture these detailed temporal structures in the EEG signals, we employed shorter temporal convolution kernels. Specifically, we reduced the kernel length of EEGNet from 64 to 6, which processes time slices of 24 ms. For the EEGInception model, we reduced the scale numbers determining the temporal filter lengths from the original 500, 250, 125 to 100, 50, 25, resulting in temporal convolutions of three shorter lengths: 25, 12, and 6. The hyperparameters were also optimized to fully explore the upper limits of the models. Specifically, for the DeepConvNet and the LMDA, the adjustment was made in the number of filters in a typical convolution operation. For the EEGNet, the adjustment was made in the number of filters in each separate spatial convolution. For the EEGInception, the adjustment was made in the number of temporal filters in each branch of Inception module. For the DeepConvNet, we also modified the number of filters in the following convolutional blocks. Finally, the dropout rate was also fine-tuned. These modifications were all found to significantly enhance the classification performance in terms of validation accuracy, i.e., without overfitting issues. All of the architecture end at a final flat layer with output units of 26, corresponding to the size of English alphabet, to fit the number of classes in our decoding task.

The four neural network models were trained and evaluated using a 10-fold cross-validation procedure on each subject. Specifically, the 300 samples from each class were partitioned into 10 equally-sized bins. In each training and testing session, one bin is held out for validation, while the other nine bins were used to train the models. Each bin was used as validation set exactly once, resulting in a total of 10 training sessions. In each session, the training was stopped when the model performance was not able to improve. The highest validation accuracies attained in each fold were recorded for subsequent statistical analysis. To mitigate any systematic structures in the EEG dataset, the samples in each class were shuffled prior to binning.

## 3. Results
### 3.1. Similarity matrices revealing letter specificity

The method for calculating the similarity of signals from every two letters is described in Section 2.2. Fig. 2(a) presents the similarity matrix for Subject 1, calculated from the dataset used in machine learning classification with optimal hyperparameters (time window: 0–1600 ms; filtering: 0.1–45 Hz). As expected, the similarity matrix has a clearly visible diagonal, demonstrating within-letter similarity and cross-letter dissimilarity in the dataset. This provides solid evidence for the feasibility of machine learning classification. To examine the robustness of the diagonal pattern (which indicates letter specific EEG features), we further calculated the similarity matrices across the combinations of different time windows and frequency bands. Fig. 2(b) presents the results for the following time epochs: 0–1000, 100–1200, 200–1400, and 300–1600 ms, filtered by bandpass with the following frequency ranges: 1–10, 5–20, and 9–30 Hz. We observed a visible diagonal in each of the matrices, indicating high robustness in letter specificity across a wide range of hyperparameters. For the architecture of the EEGInception, such a consistent letter specificity in different frequency components could support the usage of temporal filtering of different sizes, which aims at extracting both fast- and slowly-varying features in the data. Moreover, it is noteworthy that the lower frequency components (e.g., 1–10 Hz) generally demonstrated stronger letter specificity than the higher ones (e.g., 9–30 Hz), with a more conspicuous diagonal in the similarity matrix.

In Fig. 2(c), we present the similarity matrices for the other subjects. The data from Subject 5 also produced a visible diagonal on the similarity matrix, while for Subject 2 and 3 this pattern is weaker but still identifiable, and for Subject 4 such a pattern is hardly observable. This shows a high variability of the letter specificity patterns across individual participants. However, the absence of a conspicuous diagonal does not negate the possibility of a classification because the classification as will be implemented by the complex neural networks may extract higher-level information or features. The similarity measurement here is sufficient but not necessary for the success of machine learning classification. The letter specificity might not be well illuminated by Pearson correlation which is a measurement for linear similarity, and a neural network model could extract more complicated letter-specific patterns from the datasets. However, a machine learning model is more likely to achieve excellent classification performance if the dataset displays strong letter specificity in the similarity matrix.

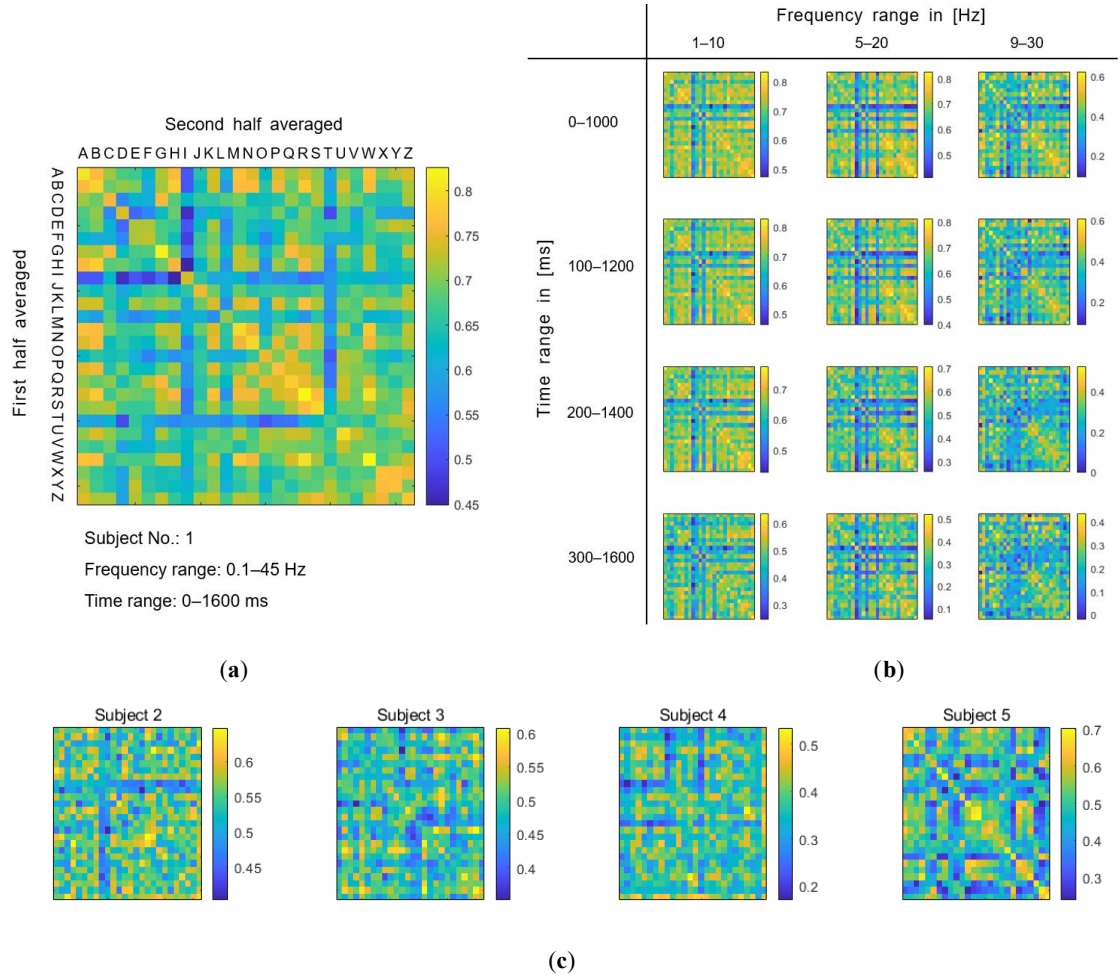

Fig. 2. Similarity matrices indicating letter specificity in EEG features. (**a**) Similarity matrix of the dataset of Subject 1 used for machine learning classification. The time range is 0–1600 ms, and a bandpass filter of 0.1–45 Hz was applied. (**b**) Similarity matrices of different combinations of time and frequency ranges for Subject 1. (**c**) Similarity matrices for the other participants, adjusted to the same color range. These similarity matrices are based on signals used for machine learning classification (time range: 0–1600 ms, frequency range: 0.1–45 Hz). The visible diagonals indicate within-letter similarity and cross-letter dissimilarity in the dataset.

**3.2. Visualization of the average patterns**

To exemplify the same-letter similarity and different-letter dissimilarity, we visualized the average patterns of two different letters following the methods described in Section 2.3. Specifically, we calculated the two average patterns of G and I, and then visualized the spatial distribution of their potential values at the following time steps: 300, 500, 700, 900, 1100, and 1300 ms (Fig. 3, upper right). This allows us to intuitively compare the scalp distributions between the average patterns of the same and different letters. Notably, we observe relatively consistent spatial distributions in each pair of average signals from the same letter: the two average patterns of G (first and second rows) or I (last two rows) present similar scalp distributions over time course. Additionally, the difference between scalp distributions from two different letters is visually identifiable. For example, the first average of G (first row) is more similar to the second average of G (second row) than to the first average of I (third row) at each time step. This same-letter similarity and different-letter dissimilarity in scalp distribution suggest that spatial information is important for distinguishing EEG patterns of different letters, providing a foundation for the use of spatial filtering in machine learning classifiers.

Additionally, we applied the principal component analysis (PCA) to each of the four average patterns (two Gs and two Is). The scalp recordings from the 24 channels at each time step were treated as one observation, producing principal components with the same number of time points as the signals. The first few principal components would then reveal the governing temporal dynamics underlying the signals. In the bottom panel of Fig. 3, the first three principal components (PCs) of each average pattern are displayed. As expected, the PCs of two average patterns for the same letter are closer to each other than the PCs of two patterns for different letters. Across all the orders of PCs, the overlapping of two PCs from the same letter and the separation of PCs from different letters are visually observable. This letter-specific pattern in temporal dynamics provides a foundation for the use of temporal convolution in the neural network models.

However, the patterns displayed in Fig. 3 also indicate the limitations in the data quality and the challenges for machine learning classification. Note that while the average patterns of the same letter are more similar than those of two different letters, they are still far from being identical, as shown by the scalp distributions at multiple time points. Even though the patterns were averaged from 150 samples, this did not produce highly stable event-related potentials (ERPs). In fact, the ERPs for each English letter could vary across different experimental sessions and even different periods within the same session (note that the participants came to the lab on different days, and the samples were generated in multiple separate periods). To eliminate any systematic change across trials, we have shuffled the samples from each letter before calculating the average patterns, but we observed that different random shuffling still produced very different average patterns for the same letter. This suggests that the individual samples may present much less letter-specific patterns due to varying signal-to-noise ratio, which poses significant challenges for machine learning classification.

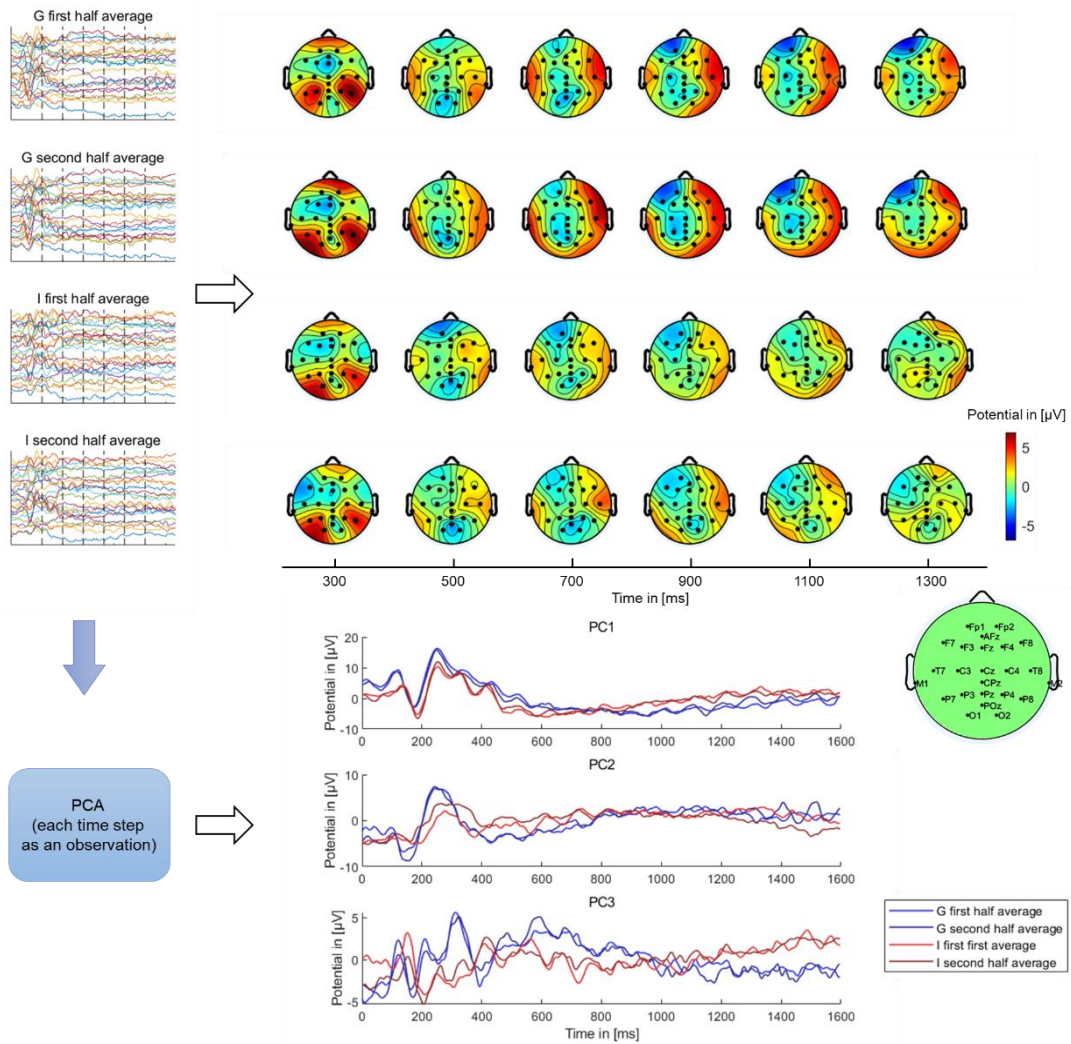

Fig. 3. The principal components and scalp distributions of the average patterns of letter G and I. Upper left: illustration of the four average patterns, each consisting of scalp recordings from the 24 channels. The vertical dashed lines indicate the time points for visualizing scalp distributions. Upper right: the spatial distributions of the scalp recordings of each average pattern at 300, 500, 700, 900, 1100, and 1300 ms. The additional topographical map shows the channel labels and distributions. Bottom: the first three principal components (PCs) of each average pattern. Each time step of the average patterns was regarded as an observation when conducting PCA, producing PCs reflecting the governing temporal dynamics.

### 3.3. Classification accuracy

As detailed in Section 2.4, the four neural network models were evaluated following a 10-fold cross-validation procedure on each participant. For the dataset from each participant, the models were trained in 10 folds on different dataset splits, yielding 10 highest validation accuracies for each model.

To facilitate cross-model and cross-subject comparisons, we present the 10 highest validation accuracies for each model and subject using a box plot in Fig. 4. The full list of validation accuracies can be found in Supplementary Table 1.

In general, the four models exhibited a moderate level of performance on the scalp-recorded EEG data, with validation accuracies showing a notable degree of cross-subject variability. The models demonstrated the highest potential to achieve a validation accuracy of up to 20% on Subject 1 and 5 but compromised accuracy on Subject 2, 3, and 4.

To highlight the upper limit of these models in decoding handwriting imagery, we presented the maximums of the 10 highest validation accuracies achieved by each model trained on each subject in Table 1. Notably, the DeepConvNet attained the highest accuracy of 22.44% on Subject 1, while the EEGInception achieved 22.31% on Subject 5 – the best performance among the models on those respective subjects. Although these classification accuracies may not be impressively high, they significantly exceed the chance level of 3.85%, indicating the potential of decoding handwriting imagery from scalp-recorded EEG signals. It is also noteworthy that even the lower-performing subjects (2, 3, and 4) exhibited accuracies exceeding the chance level, and this suggests the presence of letter-specific patterns in the EEG data across all subjects.

Table 1. The highest validation accuracies achieved by each model on each subject over the 10 folds.

| Subject | 1 | 2 | 3 | 4 | 5 |
|---|---|---|---|---|---|
| DeepConvNet | 22.4359 | 12.0513 | 9.3590 | 8.9744 | 17.9487 |
| EEGNet | 16.9231 | 9.3590 | 8.9744 | 8.8462 | 22.0513 |
| EEGInception | 22.3077 | 9.7436 | 7.8205 | 9.6154 | 22.3077 |
| LMDA | 18.2051 | 8.5897 | 7.8205 | 8.3333 | 18.0769 |

The classification performance of each model varied across subjects, and it is noteworthy that this cross-subject variation in the classification performance aligns with the results of the descriptive analysis. Recall that Subject 1 and 5 displayed conspicuous diagonals in their similarity matrices (Fig. 2), which indicates strong letter specificity for ML classification. This corresponds to the generally higher classification accuracies of Subject 1 and 5 compared to Subject 2 through 4 (Fig. 4). However, the accuracies for the latter group of subjects still exceeded the chance level, suggesting that the neural networks were able to extract letter-specific patterns from all subjects. This confirmed our assertion that the visible diagonal of the similarity matrix indicates the feasibility of classification but is not a necessary condition for it. The causes of this cross-subject variation will be discussed in Discussion section.

Variations in classification performance can be observed across the four models. A one-way ANOVA test ($\alpha = 0.05$) based on the 10 accuracies of each model revealed statistically significant differences in the classification performance across the four models for Subject 1, 2, 3, and 5 ($p < 0.05$), but not for Subject 4 ($p > 0.05$). Additionally, the relative performance of different models sometimes varied or even reversed across different subjects. For example, the DeepConvNet outperformed EEGNet on Subject 1 ($p < 0.05$, one-tailed $t$-test), but EEGNet outperformed DeepConvNet on Subject 5 ($p < 0.05$, one-tailed $t$-test). It is also noteworthy that the LMDA generally did not outperform other models, with its average accuracy lower than the other models on Subject 2 through 5, and only exceeding the EEGNet on Subject 1. Possible reasons and potential improvements will be discussed in the Discussion section.

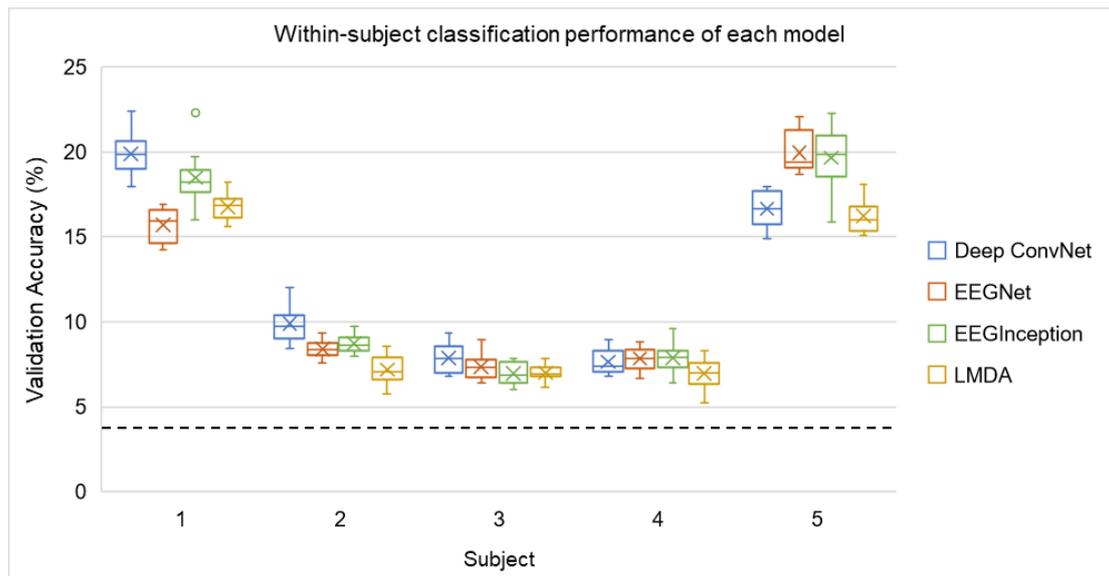

Fig. 4. Classification performance of each neural network model trained on each subject. The neural networks were trained and validated using a 10-fold cross-validation method. The 10 highest validation accuracies of the corresponding models and subjects are displayed here in a box plot. The dash line indicates the chance-level accuracy 3.85% (1÷26×100%).

## 4. Discussion

This work explored the potential of using non-invasive EEG data for brain-to-text BCI applications. We trained four different state-of-the-art CNN models on our dataset from five participants and demonstrated the capability of decoding non-invasively recorded EEG signals into the content of handwriting imagery. Although the accuracy of 20% (compared to 3.8% chance level) from some participants is not impressively high, this work is the first one that lays down the foundation of non-invasive EEG-based brain-to-text BCI. In the following, we will discuss several important issues in this study and provide recommendations for research endeavors.

It should be noted that our dataset has a limited sample size, which includes 7,800 EEG data samples (300 samples for each letter) for each participant. And because of the significant cross-subject variability, we did not attempt to explore the potential of further improving the accuracy after aggregating all participants' data (but will be explored in future works). As compared to a typical deep learning application, the available data in the current work are far from being sufficient for exploiting the power of deep neural network models. For instance, in a typical image classification task, a deep neural network may be trained from hundreds of thousands of image samples per category. Inadequate sample size hinders the model's ability to learn the multi-level features from the variability of data. As such, the accuracy level reached in the current study does not fully reveal the true potential of using non-invasive EEG to decode handwritten content. Furthermore, since the participants visited the lab on multiple occasions, there might be variations across different visits, further compromising the consistency of data in certain ways.

The analysis presented in Fig. 4 (Section 3.3) reveals a notable degree of cross-individual variation in the performance of the model. This subject-to-subject variability reveals another aspect of complexity in EEG signals and highlights the challenges in developing a one-size-fits-all classifier for different individuals. The subject-to-subject variability may arise from factors such as variations in the signal-to-noise ratio in different participants, which could be attributed to task-related factors such as participants' level of attention and muscular movements, or task-unrelated factors such as the overall quality of EEG data produced by different participants (e.g., due to skin conductivity). To refine and optimize the model's performance in future research, it is imperative to conduct investigations into the underlying factors contributing to subject-specific variations in accuracy. One direction is to develop individually specific hyperparameters and other model settings that fully exploit the potential of each individual by considering individual specificity.

Another factor that could substantially affect the accuracy result is the pre-processing steps in EEG data analysis. The current preprocessing procedures employed in our study, including filtering, epoching, and normalization, are simplistic and may not effectively eliminate all irrelevant components that may hamper the learning process of the machine learning procedure. It is important to recognize that scalp-recorded EEG signals are susceptible to various sources of noise, such as eye blinks, muscular movements, and electrode artifacts, all of which can interfere with the components relevant to handwriting imagery. Therefore, developing a set of pre-processing steps that are specifically tailored for handwriting imagery data may be another direction that could further improve the accuracy. Additionally, to echo the previous point about individual differences: it can be also considered to apply individually specific pre-processing steps on each participant in the context of non-invasive EEG-based decoding.

The current neural network architectures employed for handwriting imagery EEG classification may also have further potential to improve. While our CNN architectures in this study appeared to be able to capture both temporal and spatial relations in the EEG data, which are fundamental features of EEG signals, it predominantly utilizes local information, such as neighboring time points or electrode channels with consecutive labels. There could be other valuable information for decoding the imagined contents that has not been revealed by the current network models. For example, in this work, we have not yet systematically exploited features in the data that are specifically related to handwriting, such as the direction and turning points of imagined handwriting trajectories. Additionally, our analysis also did not consider the inter-relationships between different frequency components of the EEG signals (e.g., cross-frequency coupling), which could potentially provide valuable information for classification. Last but not least, the relatively low accuracies achieved by the LMDA were possibly caused by the limited number of channels in our dataset with limited spatial information, which is a key design consideration in the LMDA's channel attention module. In fact, the authors of [9] noted that when the number of channels is low, it is meaningless to incorporate the channel attention module. In sum, there could be more complex spatial information useful for decoding the imagined contents. These advanced features will be further explored in future works.

To increase the sample size in future research, we recommend recruiting more participants to undertake the same handwriting imagery task. This can be achieved by initiating cross-laboratory collaborations to leverage the benefits of a large sample size that is only available through large-scale collaborations. A larger and more diverse dataset will facilitate model training and generalization and contribute to more robust and reliable results. However, standardized protocols have to be established before cross-laboratory cooperation. Since the tasks in such handwriting imagery are relatively simplistic, standardization of task protocols would not be a major challenge.

To mitigate the subject variability issue, we recommend implementing certain measures during data acquisition. First of all, it is very important to instruct participants to strictly focus on the task and minimize movement during the handwriting imagery stage. Participants should also be instructed to standardize their handwriting style and speed during the task, even though the task is an imagery task. One of the limitations of the current task setting is that there is no additional procedure to verify if the participant is corporative or not in terms of internally generating the mental processes related to handwriting imagery. Because it is practically difficult to test the genuineness of mental processes in our current task, it is challenging to develop an objective criterion for this purpose. However, there are still potentially useful ways to improve the genuineness of the participant's task performance. One potential way to improve this aspect is to use neural feedback to enhance the alertness level of the participant. For instance, the ERPs from each block can be generated in real time and the signal-to-ratio can be immediately calculated. The participant's objective is to improve the signal-to-ratio over blocks. In this setting, even if the signal-to-ratio is not necessarily easy to improve using mental power, the attention or motivation of the participants could be improved as compared to the current version without any neural feedback.

The preprocessing stage could be further improved to possibly further push up the upper limit of the classification accuracy. The current preprocessing procedure is primarily based on a relatively standardized procedure developed from basic cognitive research based on EEG and ERP. They may be more suitable for analyzing subtle effects in the ERP data, but may not be ideal procedures for refining EEG data from motor imagery paradigms. One direction is to incorporate more advanced time series analysis techniques to develop complex filters for the raw EEG data. The underlying structure patterns of noise in EEG data should be understood in greater detail, particularly, how the noise is differentiated from the imagery-related signal components. One theoretical proposal is to consider using the pre-stimulus time slice as a baseline to develop complex spatiotemporal filters to enhance the extraction of signal components in the post-stimulus time window that is specific to handwriting imagery.

Regarding the neural network architecture, here we also propose several avenues for future improvement. Firstly, researchers should delve deeper into the intrinsic characteristics of the EEG data relevant to handwriting imagery trajectories, and use those imagery-specific features to inform the development and design of feature extraction algorithms. This can be achieved by exploring representations of the signals that exhibit enhanced similarity within letters and dissimilarity across letters, as shown in the inter-letter similarity matrix (Fig 2). The understanding of more discriminative features will inform the development of models more specifically oriented to capture them. Secondly, incorporating advanced algorithms and architectures designed for time series classification, such as the Self-Attention mechanism and the Transformer model, should be considered. These techniques, which have gained prominence in natural language processing, have the capability to capture complex temporal relationships, including long-distance dependencies, which could be more informative for classification tasks. However, the caveat is that those models usually require enormous amounts of data for training. Lastly, to leverage the information contained in different frequency components over time, conducting time-frequency analysis and utilizing the internal relationships in the time-frequency plot, such as cross-frequency coupling, would be beneficial. Also, to fully utilize the spatial relation, it would be advised to increase the number of channels of EEG recording. A larger number of channels would be advantageous to architectures that specifically consider spatial relations, like the LMDA.

In addition to the aforementioned suggestions for improving letter classification, the application of the autospell technique holds the potential for further boosting the practicality of non-invasive EEG-to-text decoding. Autospell correction is a widely utilized technique in various applications, such as word processors and presentation software, enabling automatic correction of misspelled words based on established language models. Therefore, if the potential of the autospell technique is fully exploited, we only need to use the information from non-invasive EEG as an axillary channel to guide the language generation processes. The availability of autospell would allow the EEG-to-text decoding to stay at a medium accuracy level, e.g., around 50%, while being able to support feasible real-time translations. Given the noise level in non-invasive EEG data, reaching a level of close-to-100% accuracy may be too ambitious. However, our current result has achieved a level that is more than five times the chance level, which makes reaching a medium level of accuracy a possible objective in the near future.

Finally, another limitation of the current study lies in the task paradigm design. In our design, the participants were instructed to start imagining the handwriting of the letters after seeing the visual presentation of the letters on the screen. We understand that, technically, the visual content processing could partially contribute to the within-letter specificity and cross-letter differences. Although this is technically the case, here we provide arguments that this effect does not account for the full effect and the reasons why we made this design. In traditional ERP research, normally, having simple slight visual change without cognitive implications usually leads to very subtle changes in the ERP [15][26]. Even if we assume that the sensory processing partially contributed to the effect, the similarities matrix has shown that the effect persists to the later time window after 300 ms (after which the sensory processing is normally considered to have finished). While we acknowledge the technical possibility of contamination from sensory processing, one of the main reasons for the current design (rather than a delayed response) was to increase the sample size (by shortening the inter-stimulus interval) for training. An alternative design is to delay the imagery process by adding a universal cue [24]. However, there could still be issues such as incorporation from the participants. Besides, holding the character in mind, although not part of sensory processing, could still contaminate the handwriting imagery process in principle. Taken together, we note the limitation of the task paradigm design (which was a compromise for achieving a larger size) and recommend making improvements in this respect in the future.

## 5. Conclusions
In this study, we first explored the feasibility of classifying non-invasively recorded EEG of handwriting imagery for a brain-to-text BCI. Using a diversity of cutting-edge convolutional neural networks for EEG classification, we established a baseline for future research in this domain. The use of non-invasive EEG recording techniques in our study offers advantages in terms of accessibility and usability. While the model's performance for an applicable BCI remains moderate, the achieved validation accuracy is noteworthy compared to random letter selection, indicating the potential for accurate translation of brain signals associated with handwriting imagery into characters using non-invasive EEG. Moving forward, future research endeavors will focus on enhancing decoding accuracy with more sophisticated network architectures and meticulous preprocessing techniques. These efforts aim to unlock the full potential of non-invasive handwriting imagery EEG for robust and reliable brain-to-text BCI systems.

**Author contributions**

Hao Yang contributed to EEG data analysis, machine learning classification, and manuscript drafting. He was responsible for the overall framework and writing of the manuscript. He conducted descriptive analysis on the EEG data as outlined in this manuscript and made significant contributions to the classification section by exploring various network architectures, training the four CNNs on the dataset, organizing the training results, and performing statistical comparison between the CNNs.

Guang Ouyang's contributions included experimental design, data collection, and manuscript preparation. He designed the experimental paradigm for handwriting imagery EEG and collecting the data utilized in this study. Additionally, he actively participated in editing the manuscript, making substantial revisions to the improve clarity and overall composition.

## Data availability

The preprocessed EEG data that support this study are openly available at this URL: https://github.com/guangouyang/handwritingBCI.

## Funding

This work was supported in part by the Seed Fund for Basic Research from the University of Hong Kong under Grant/Award 2302101550 and in part by Hong Kong Research Grant Council under Grant/Award 17609321.